# Coordinated, Interactive Data Visualization for Neutron Scattering Data

D. J. Mikkelson[2], R. L. Mikkelson[2], T. G. Worlton[1], A. Chatterjee[1], J. P. Hammonds[1],
P. F. Peterson[1], A. J. Schultz[1]

[1]*Argonne National Laboratory, Argonne, IL 60439, USA*
[2]*University of Wisconsin-Stout, Menomonie, WI 54751*

ABSTRACT

The overall design of the Integrated Spectral Analysis Workbench (ISAW), being developed at Argonne, provides for an extensible, highly interactive, collaborating set of viewers for neutron scattering data. Large arbitrary collections of spectra from multiple detectors can be viewed as an image, a scrolled list of individual graphs, or using a 3D representation of the instrument showing the detector positions. Data from an area detector can be displayed using a contour or intensity map as well as an interactive table. Selected spectra can be displayed in tables or on a conventional graph. A unique characteristic of these viewers is their interactivity and coordination. The position *pointed at* by the user in one viewer is sent to other viewers of the same DataSet so they can track the position and display relevant information. Specialized viewers for single crystal neutron diffractometers are being developed. A "proof-of-concept" viewer that directly displays the 3D reciprocal lattice from a complete series of runs on a single crystal diffractometer has been implemented.

INTRODUCTION

The Integrated Spectral Analysis Workbench (ISAW) is an object oriented modular system being developed at IPNS to visualize, reduce and analyze data from neutron scattering experiments. ISAW is implemented in Java for portability and is organized around several fundamental unifying concepts: Data blocks, DataSets, Operators, DataSetViewers, Retrievers and Writers. After a brief description of these concepts, this paper will describe several unique aspects of the viewers in more detail.

A Data block is a software object representing a function or histogram of one variable, *x*, on an interval. The Data block may contain a table of measured *y* values, a list of events, or an expression or function to calculate the values as needed. For time-of-flight neutron scattering experiments, a Data block typically stores the histogram of counts vs. time-of-flight. In addition, the Data block contains a list of named attributes that are used to specify meta-data such as detector size, position, initial flight path length, incident energy, etc. Error estimates may also be included in the Data block.

A DataSet is a collection of Data blocks, together with units, labels and a list of attributes that apply to the entire DataSet. All Data blocks in a DataSet must have the same units, however, Individual Data blocks may cover different intervals. Each DataSet includes a list of Operators that are appropriate for that DataSet. The DataSet also records information about a current region of interest in the form of a set of selected Data blocks and a selected sub-interval. Finally, the DataSet keeps track of what Data block and x-value is being *pointed at* by the user.

A DataSet can be loaded using a Retriever object, saved using a Writer object, operated on by Operator objects, and viewed using one of several DataSetViewer objects, under the control of a ViewManager.

CURRENT VIEWERS

Currently, six viewers have been implemented:

-Image View: Displays the values of a collection of Data blocks as rows in an image. Selected Data blocks and the Data block that is currently pointed at are displayed as line graphs across the bottom of the viewer (Mikkelson 1995).

- Scrolled Graph: Displays all of the Data blocks as ordinary line graphs in a large scrolled pane.

- Selected Graph: Displays the selected Data blocks as ordinary line graphs in one combined graph. This is a static graph of the data with calibrated axes.

- 3D: Displays the y value for each Data block at one x value. The y value is displayed as a colored rectangle in 3D at the detector position. The x values can be changed interactively and controls are provided to automatically step through a sequence of x values, producing a "movie".

- Table: Displays a table of values from the Data blocks, together with meta-data as requested by the user. Several basic tables are provided for convenience.

- Contour: Displays an image or contour map similar to the 3D view, but in a 2D plane.

The Contour view and Selected Graph view are built using the Scientific Graphics Toolkit from NOAA. (http://www.epic.noaa.gov/java/sgt) All of the other viewers were built from the graphics primitives available in Java. Figures 1, 2, and 3 give several views. Figure 4 gives a composite with the three other viewers.

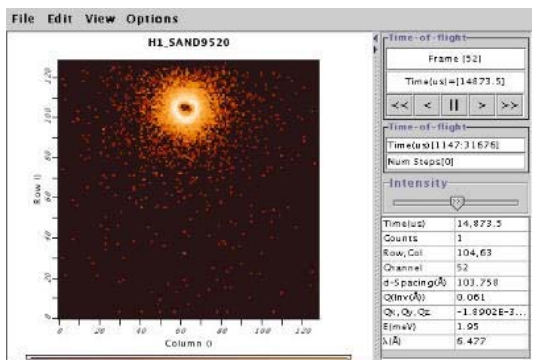

**Figure 1.** Intensity plot in the Contour viewer of one time slice of a data set from the SAND instrument at IPNS. The controls at the right allow stepping to different time slices, rebinning the data in time-of-flight, and adjusting the color scale. The display in the bottom right corner shows physically meaningful data about any point on the display that the user selects.

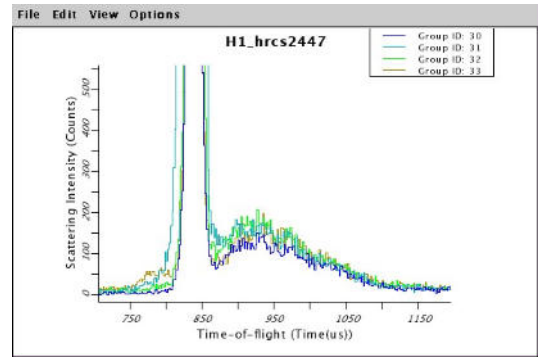

**Figure 2:** Selected graph view of four spectra from the HRMECS instrument at IPNS.

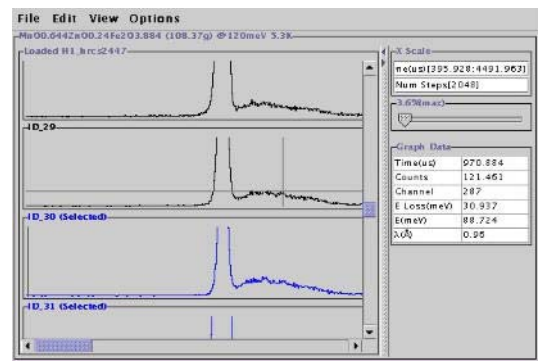

**Figure 3:** Scrolled graph view of several spectra from the HRMECS instrument at IPNS. All spectra in the DataSet can be rapidly reviewed by sliding the knob on the scroll pane. The controls on the right allow rebinning in time and setting the vertical scale factor for the graphs.

These viewers have several powerful features. First, they are not just fixed displays, but include many options and interactive features. For example, most of the viewers include a table of information about the point on the viewer that the user is pointing at. This table is configured automatically based on the conversion and information operators that are included with the DataSet. Typical time-of-flight DataSets include Operators to convert the DataSet to wavelength, d-spacing, momentum transfer, etc. Since the Data blocks include the necessary meta-data these calculations can be easily performed.

Additional controls are included with different viewers, as appropriate. For example, the 3D viewer includes controls to adjust the point of view, and an animation controller widget that lets the user step through different time slices one slice at a time, or to step through the time slices automatically like a movie. Other controls allow adjusting the color scales and scaling for the line graphs that are part of the image and scrolled graph views.

In addition to the interactive capabilities of the individual viewers, the viewers can also work together. The collaboration of the viewers is implemented using the Observable-Observer design pattern(Coad 1999). This design pattern works as follows. The observable object maintains a list of objects that observe it. When the observable object is altered, an update message is sent to all of its observers. When the observers receive the update message, they can get additional information from the observed object and respond as needed. This design pattern preserves the independence of the objects. An observer does not even need to know about the existence of any other observer. Similarly, the observed object does not need to know any details about its observers since it only sends the update message to it's observers.

For coordinating several viewers of one DataSet, the DataSet is the observable object. The viewers for that DataSet are the observers. When the user points at a position in a view of the DataSet, the viewer sets the *pointed at* Data block and the *pointed at* x-value in the DataSet. It then asks the DataSet to notify its observers that the *pointed at* values have been changed. The DataSet then merely sends out a *pointed at* changed message to its observers. When the viewers receive the *pointed at* changed message, they can get the new *pointed at* x-value and Data block and can update the view as needed. Examples of this are shown in Figures 4 and 5.

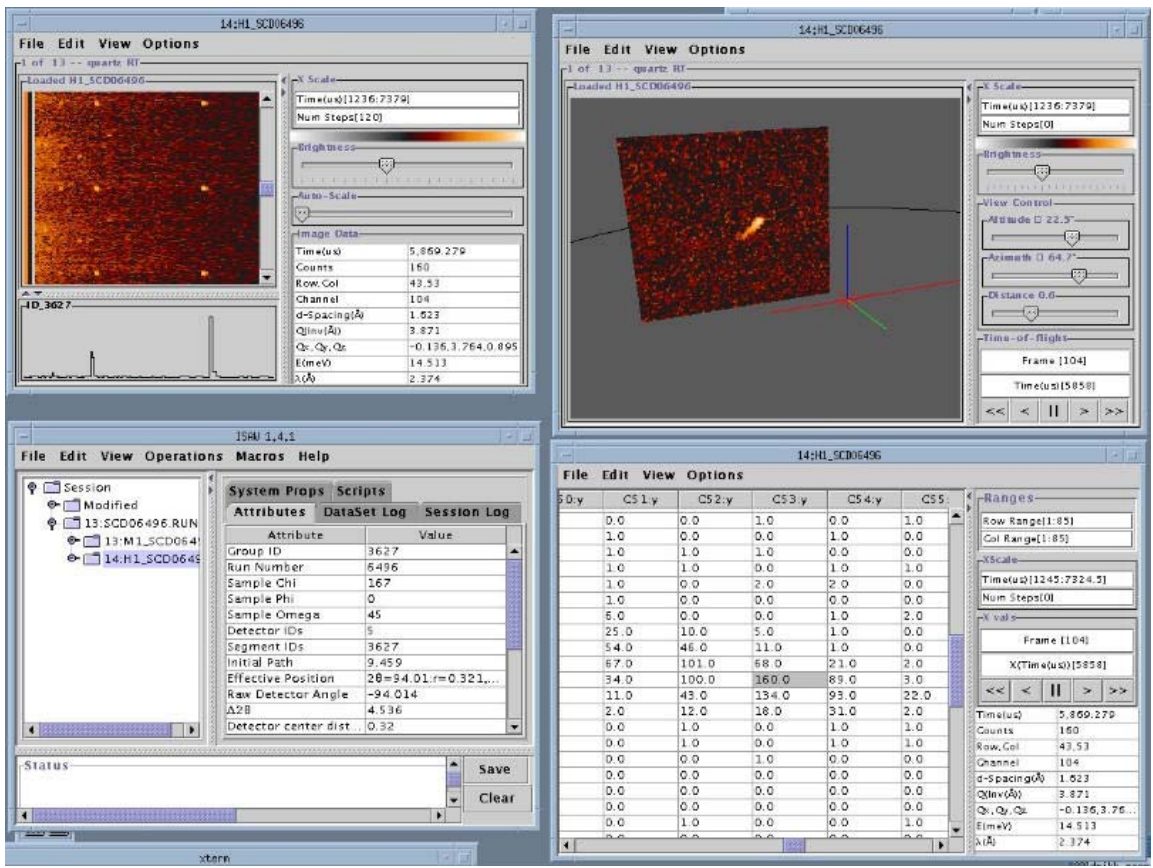

**Figure 4**: Screen shot of the main ISAW control panel with an image view, 3D view, and table view (clockwise from lower left). As the user moves the cursor to peaks on the image in the upper left, the 3D view and table view automatically switch to the same time slice and pixel. This makes it easy to find peaks in the large block of data.

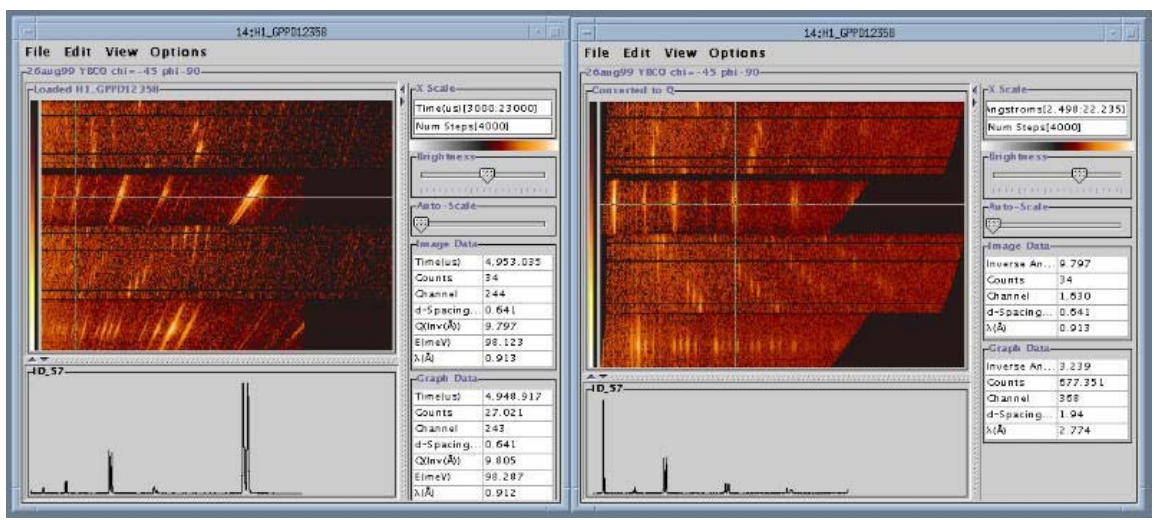

**Figure 5:** Two image views of the same DataSet from the GPPD at IPNS. The data is shown as a function of momentum transfer in the viewer on the right. As the user points at peaks in the time-of-flight spectra image on the left, the cursor in the momentum transfer image moves to the same peak. The pointed at spectrum appears as a graph versus time-of-flight at the bottom of the image viewer on the left and versus momentum transfer at the bottom of the image viewer on the right. Controls for each viewer allow rebinning, adjusting the color scale, and the vertical scale factor for the line graph.

FUTURE PLANS

Since the DataSet contains the relevant meta-data, there are many possibilities for further development of viewers. Some preliminary work has been done on viewers for data from single crystal diffractometers. Figures 6, and 7 show several views of the reciprocal lattice for quartz, obtained from a set of 13 runs on the SCD at IPNS. The first of these is not aligned with any crystal plane. The other one is aligned with one of the crystal planes. These images were obtained as follows:

1. For each DataSet, the histogram bins that exceeded a specified threshold were transformed to the corresponding momentum transfer vector (Q).

2. The goniometer angles chi, phi and omega were used to map the Q vectors to a common frame of reference.

3. For each such Q vector, a 3D point with a color corresponding to the number of counts in the histogram bin was placed at the calculated Q position.

In this preliminary form, the viewer only allows moving around the 3D display and zooming in on particular portions of the reciprocal lattice.

Planned enhancements will allow the user to interactively measure distances between points and/or planes, fit vectors and planes to the data, and select planes for generating contour plots or slices through the data.

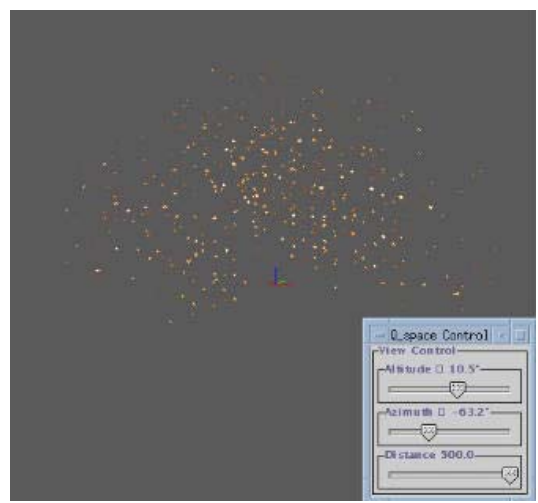

**Figure 6:** Reciprocal lattice for quartz obtained by combining data from 13 different crystal orientations. The control allows zooming in and adjusting the position of the observer relative to the center of the lattice. In this figure the position of the observer is not aligned with any crystal axis.

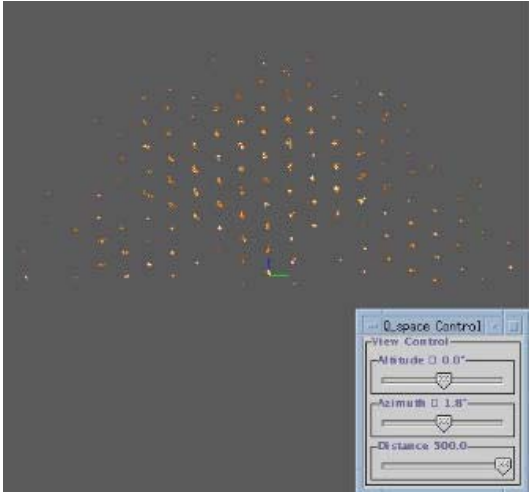

**Figure 7:** Reciprocal lattice for quartz using the same data as in Figure 6.  In this figure the position of the observer is aligned with one of the crystal axes showing the hexagonal structure in the reciprocal lattice.

ACKNOWLEDGMENT

This work was supported by the Intense Pulsed Neutron Source division and the Division of Educational programs of Argonne National Laboratory, and by the National Science Foundation, award number DMR-0218882. Argonne National Laboratory is funded by the U.S. Department of Energy, BES-Materials Science, under Contract W-31-109-ENG-38.